\documentclass[fleqn,twoside]{article}
\usepackage{espcrc2}
                                                                                
\usepackage{graphicx}

\newcommand{\AmS}{{\protect\the\textfont2
  A\kern-.1667em\lower.5ex\hbox{M}\kern-.125emS}}
\title{
Nonperturbative Renormalization for Domain Wall Fermions
and the Chiral Condensate
 }
                                                                                
\author{Azusa Yamaguchi
\address{Department of Physics,
        Columbia University,New York, USA} }

\begin{document}
                                                                                
\begin{abstract}
We study the chiral condensate, $\langle\bar\psi\psi\rangle$,
and various quark bilinear vertex functions for domain wall
fermions at different lattice scales, with both the Wilson and
DBW2 gauge actions, in both quenched and dynamical fermion
simulations.  We use the vertex functions to calculate renormalization
factors within a non-perturbative scheme.
\end{abstract}
                                                                                
\maketitle
\section{Introduction}
\begin{figure*}
\vspace{-1.0cm}
\begin{tabular}{cc}
\includegraphics[width=18pc,height=10pc]{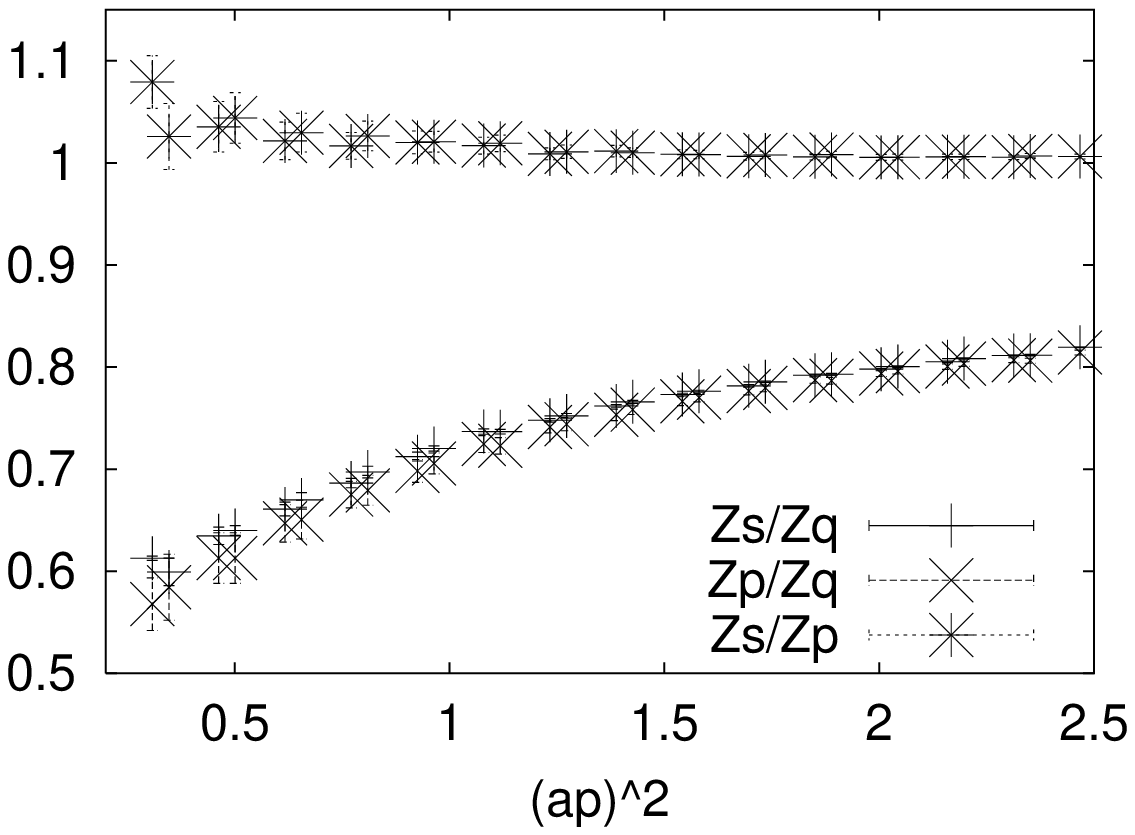}
\vspace{-0.6cm} &
\includegraphics[width=18pc,height=10pc]{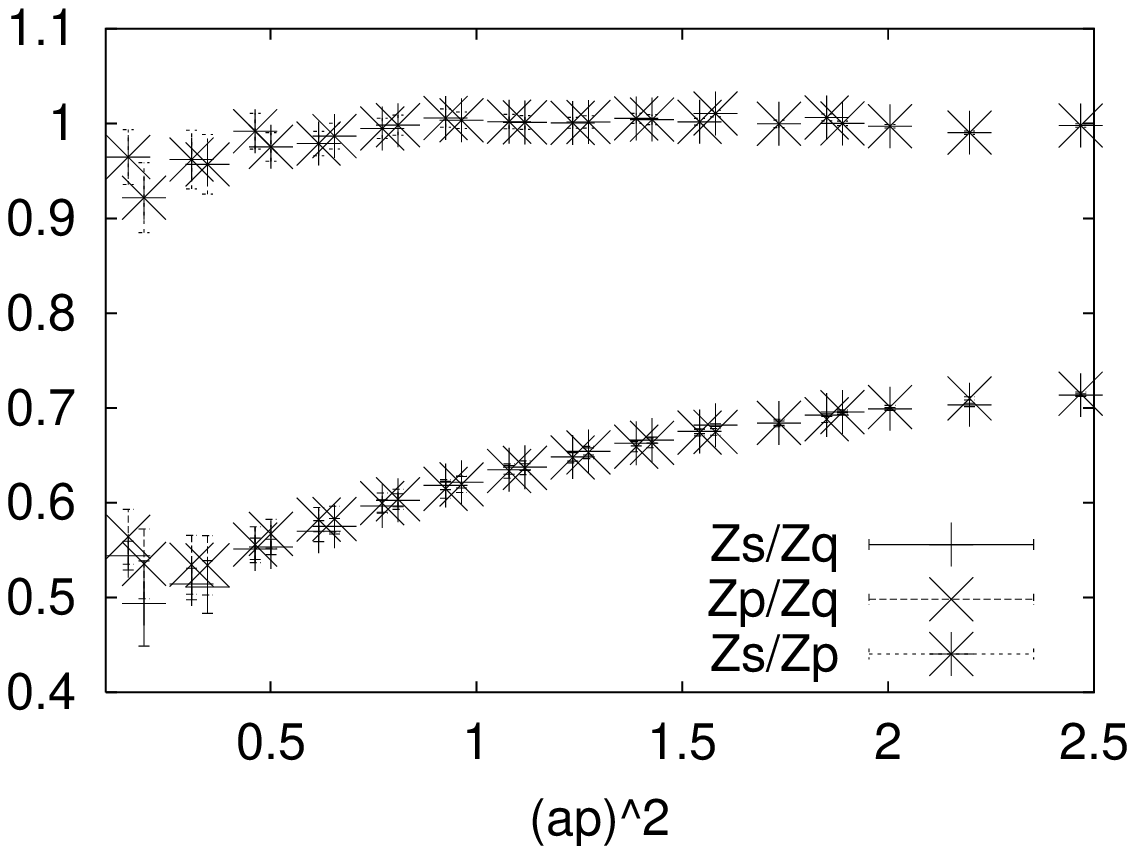}
\end{tabular}
\caption{$Z_S/Z_q$ and $Z_P/Z_q$ plotted versus $p^2$ for 
$a^{-1}=3$GeV(left) and dynamical fermions(right).}\label{fig:ZSP}
\end{figure*}
The RBC collaboration has performed simulations with domain wall fermions (DWF) and the
DBW2 gauge action and presented results for weak matrix elements.
This report focuses on the chiral condensate $\langle\bar\psi\psi\rangle$
and exploits 
non-perturbative renormalization calculations to get physical values for
both quenched and dynamical fermions.
The simulation details are given the Table~\ref{Tab:sim}.
Two lattice volumes are used, $24^3 \times 48$ for $a^{-1} = 3$GeV,
and $16^3 \times 32$ for the others.  Domain wall fermion 
are used for all simulations
with three values of $L_s$: $16$ for the Wilson gauge action
and for the DBW2 gauge action with $a^{-1}= 2$GeV and 1.3GeV, $10$ 
for $a^{-1} = 3$GeV and $12$ for dynamical fermions.
Among these, data and simulation details for the Wilson gauge action 
and DBW2 gauge action with 1.3GeV and 2GeV have
already reported\cite{RBC_02}.  The DBW2, 3GeV quenched and dynamical fermion 
simulations with $a^{-1}=1.8$GeV are reported in this meeting by RBC\cite{RBC_03}.

\section{$\langle\bar\psi\psi\rangle$ measurement}

\begin{table}[t]
\caption{Simulation details. Here ``Q''  
means quenched while ``D'' means dynamical.  The 
3GeV simulation uses a $24^3 \times 48$ volume while a $16^3 \times 32$ 
volume is used for the others.}
\begin{tabular}{lcll}
 Gauge action& Fermions& $L_s$ & $a^{-1}$(GeV)    \\
\hline
 Wilson & Q & 16& $1.922(40)$ \\
 DBW2   & Q & 16& $1.97(4)$ \\
 DBW2   & Q & 16& $1.31(4)$\\
 DBW2   & Q & 10& $2.86(9)$\\
 DBW2   & D & 12& $1.80(7)$
\end{tabular} 
\label{Tab:sim}
\vspace{-7mm}
\end{table}

Using DWF, we can compute the value of the chiral condensate  
directly by evaluating 
$\langle\bar\psi\psi\rangle$ at a series of quark mass and extrapolating
to $m_f = -m_{res}$ (see the section 4).
For the quenched, $a^{-1}=3\mbox{GeV}$ case, 
we use 5 fermion masses: 
$0.008, 0.016, 0.024, 0.032, 0.04$ and $m_{res}=9.72 \cdot 10^{-5}$\cite{RBC_03}.
For the dynamical case, 
three dynamical fermion masses 
$m_f^{dyn} = 0.02, 0.03 \mbox{ and } 0.04$, 
and 5 valence masses,$m_f^{val}= 0.01, 0.02, 0.03, 0.04 \mbox{ and } 0.05$
are used.
In the dynamical case, in order to get a physical value, 
we extrapolate along the line $m_f^{val}=m_f^{dyn} \rightarrow -m_{\rm res}$.
In the dynamical simulation, 
$m_{res}=1.36 \cdot 10^{-3}$~\cite{RBC_03}.
Besides the direct measurement of $\langle\bar\psi\psi\rangle$ above,
we may also determine $\langle\bar\psi\psi\rangle$ from the
$m_f$ dependence on the pion mass,
using Gell-Mann-Oakes-Renner (GMOR) relation:
\begin{equation}
f_\pi^2 \frac{m_\pi^2}{48(m_f+m_{res})} = \langle \bar\psi\psi\rangle
\end{equation}
The values obtained using
both methods are given in lattice units in Tab.~\ref{Tab:PBP}.
Apart from the 3GeV case, the values from each method agree nicely.

\begin{table*}
\caption{Values of $\langle \bar\psi\psi\rangle$ from the GMOR relation,
$\frac{b}{48}f_\pi^2$ and from direct measurement (the last column).}
\begin{tabular}{lcllll}
Action&$a^{-1}$& $m_f$ & $b$ & $\frac{b}{48}f_\pi^2$ &$\langle \bar\psi\psi\rangle$ \\
\hline
DBW2 quench DWF&2GeV & $m_f>0.01$& 2.58(2) &$2.3(2)\cdot 10^{-4}$ & $2.0(3)\cdot10^{-4}$ \\
DBW2 quench DWF&1.3GeV & $m_f>0.02$& 5.02(41) &$1.00(2)\cdot 10^{-3}$&$9.87(5)\cdot 10^{-4}$ \\
DBW2 quench DWF&3GeV & $m_f\ge 0.008$& 1.83(3) &$9.85(13)\cdot10^{-5}$  & 1.78(2) $10^{-4}$ \\
DBW2 dynamical DWF&1.75GeV & $m_{D}\ge 0.02$& 3.70(10) &$5.12(91) \cdot 10^{-4}$ &$5.56(17)\cdot 10^{-4}$
\end{tabular} 
\label{Tab:PBP}
\vspace{-6mm}
\end{table*}

\section{Non-perturbative renormalization}

In order to get physical values for continuum observables,
results obtained from lattice simulation need to be renormalized.
The RBC group has used the non-perturbative RI renormalization
scheme which benefits from the ${\cal O}(a)$ off-shell improvement of DWF. 
A detailed explanation of this method is given in the paper~\cite{NPR}. 
There are two systematic errors in this approach: 
lattice artifacts which appear at large momenta and 
non-perturbative phenomena which appear at small momenta. 
I will describe each of these in turn.

This approach gives renormalization factors for quark bilinear operators
$\bar{u}\Gamma_i d, 
(\Gamma_i= \{1,\gamma_\nu, \gamma_5, \gamma_\nu\gamma_5,
\sigma_{\mu \nu} \})$ 
where $u$ and $d$ are quark fields, using a corresponding vertex 
amplitude $\Lambda_i$.
The renormalization factor $Z_i$ is determined by the condition:
\begin{equation}
      Z_{\Gamma_i}Z_q^{-1}
       \Lambda_{\Gamma_i,0}(p,p)|_{p^2 = \mu^2} = 1.\label{eq:ZiZq}
\end{equation}
For the case of $\Lambda_S\mbox{ and }\Lambda_P$, there are 
non-perturbative contributions which must be removed.
They are shown in the equations:
\begin{eqnarray}
  \Lambda_{P,latt}(ap,ap)&=& \frac{a^2\langle \bar{q}q\rangle}
                    {(ap)^2(m_f + m_{res})} C_1 Z_q \nonumber  \\
                         &+& Z_mZ_q \\
  \Lambda_{S,latt}(ap,ap)&=& \frac{C_1Z_q}{(ap)^2}
                      \frac{\partial a^3 \langle \bar{q}q\rangle}
                    {\partial m_f} + Z_mZ_q.
\end{eqnarray}
Subtracting the effects of the first term in each of these
equations, gives values of
$Z_S/Z_P$ close to unity over a large range of momentum, see Fig.~\ref{fig:ZSP}.

\begin{figure*}
\begin{tabular}{cc}
\includegraphics[width=18pc,height=10pc]{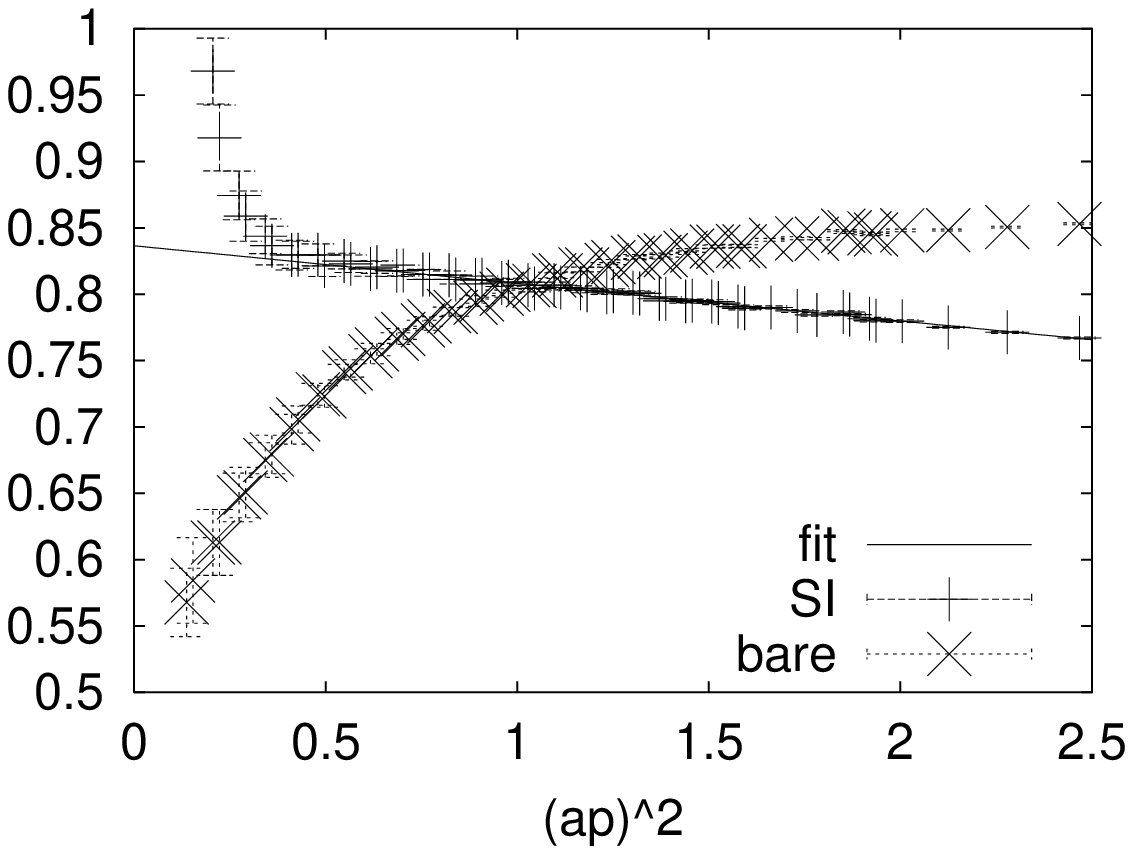}
&
\includegraphics[width=18pc,height=10pc]{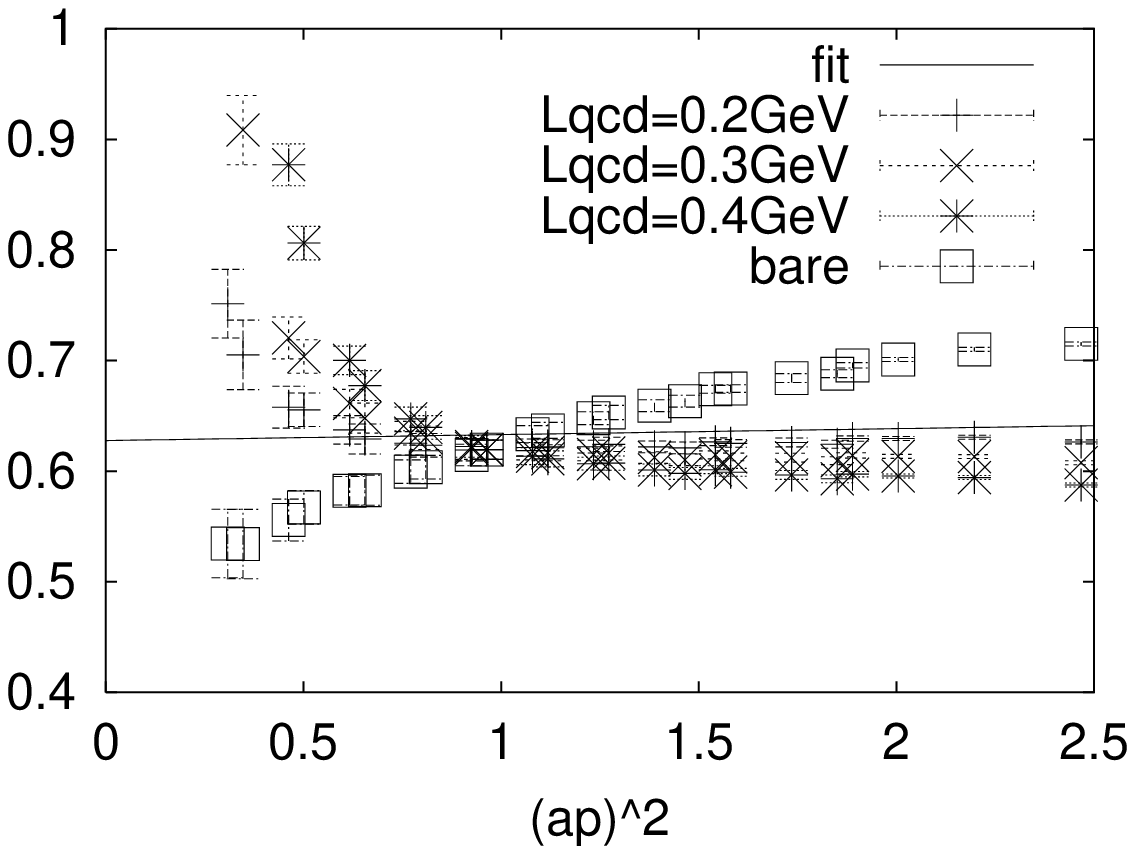}
\vspace{-0.5cm}
\end{tabular}
\caption{Bare and SI values $Z_S/Z_q$ plotted versus $p^2$ for  
$a^{-1}=3$GeV(left) and dynamical fermions(right).}
\label{fig:ZS}
\end{figure*}

\begin{table*}
\begin{center}
\caption{Results for $\langle\bar\psi\psi\rangle$ expressed in physical units}
\begin{tabular}{cccccc}
Action & $a^{-1}$ &$m_{res}$ & $Z(\overline{MS})$
& $(12\langle \bar\psi\psi\rangle)^{1/3}$ & $(12 (\frac{b}{48}f_\pi^2))^{1/3}$ \\
DBW2 quench  & $1.31(4)$  & $5.7\cdot10^{-4}$  & $0.699(16)$&$ 0.265(6)$ GeV
& $0.266(7)$GeV \\
DBW2 quench  & $1.97(4)$  & $1.7\cdot10^{-5}$  & $0.712(14)$&$ 0.235(13)$GeV
&$0.246(9)$GeV\\
DBW2 quench  & $2.86(9)$  & $9.72\cdot 10^{-5}$ & $0.832(8)$&$ 0.345(11)$GeV
&$0.282(1)$ GeV\\
DBW2 dynamical $m_D=0.02$& $1.80(7)$& $1.358\cdot 10^{-3}$& $0.469(6)$&
$0.263(10)$ GeV& $0.253(18)$GeV
\end{tabular}
\label{tab:pbp_phys}
\end{center}
\vspace{-7mm}
\end{table*}

Generally, renormalization factors in perturbation theory have logarithmic
momentum dependence.
This physical momentum dependence should be removed before 
attempting to identify ${\cal O}(a^2p^2)$ errors.
Therefore we define a scale invariant (SI)
version of the RI renormalization factor:
\begin{equation}
 \Lambda_{\Gamma_i}^{SI}((ap)^2) = \Lambda_{\Gamma_i}((ap)^2)/C_{\Gamma_i}((ap)^2)
\end{equation}
where $C_{\Gamma_i}$ is calculated through three loops and is normalized so that $C_{\Gamma_i}(\mu^2)=1$.
As an example, Fig.~\ref{fig:ZS} shows the bare $Z_S/Z_q$ and its SI version
for 3GeV and dynamical simulations.
In this figure, the dashed line is a linear fit of 
$\Lambda_{\Gamma_i}^{SI}$ as a function of $(ap)^2$, 
permitting us to remove this large momentum lattice artifact. 

To determine the conventional physical quantities defined
in the continuum theory, it is necessary to convert
renormalization factors obtained in RI-scheme to those in the $\overline{MS}$ scheme
using a relation of the form:
\begin{equation}
 \frac{Z^{RI}}{Z^{\overline{MS}}} = 1 + \frac{\alpha_s}{4\pi}Z_0^{(1)RI}
         + \frac{\alpha_s^2}{(4\pi)^2}Z_0^{(2)RI} + \cdots.
\end{equation}\label{eq:RI_MS}

In order to extract the needed factor $Z_S\mbox{ and }Z_P$ 
from Eq.~\ref{eq:ZiZq}, we must determine $Z_q$. This is best done
indirectly by using $Z_A/Z_q$ from the vertex function of the local 
axial current and $Z_A$ obtained from comparing the local and the 
conserved axial current. This method is described in detail
in Ref.~\cite{0007038} and gives values for $Z_A$ of 
$0.8876(3)$ for $a^{-1}= 3$GeV and $0.7576(6)$ the dynamical case~\cite{RBC_03}.
Using these $Z_A$ value, $Z_A/Z_q$ and $Z_S/Z_q$ from RI-scheme and
the Eq.~\ref{eq:RI_MS}, we determine the
$Z$ factors given in Tab.~\ref{tab:Zs}.
\begin{table}[b]
\vspace{-7mm}
\caption{Quenched 3GeV and dynamical Z factors.  The
dynamical results are given for $m_f^{dyn}=0.02$.}
\begin{tabular}{lcc}
Quantity & 3GeV quench & 1.75GeV dyn. \\
\hline
RI\&SI $Z_A/Z_q$ & 0.962(9) & 0.916(6) \\
RI\&SI $Z_S/Z_q$ & 0.808(3) & 0.615(19) \\
$\overline{MS}$ $Z_A/Z_q$ & 0.964(9) & 0.921(6) \\
$\overline{MS}$ $Z_S/Z_q$ & 0.904(3) & 0.7366(98) \\
\end{tabular}
\label{tab:Zs}
\end{table}

\section{$\langle\bar\psi\psi\rangle$ values in physical units} 

Using the renormalization factors discussed in the previous section,
we obtain the values for $\langle\bar\psi\psi\rangle$ 
in physical units given in Tab.~\ref{tab:pbp_phys}.
The results for $\langle\bar\psi\psi\rangle$ agree very well except for
the direct calculation in the 3GeV case which is shifted 30\% above the others.
It is expected because of the $a$ dependence of the explicit chiral
symmetry breaking in the domain wall scheme where $\langle\bar\psi\psi\rangle$ 
can be written in physical unit as,
\begin{equation}
\langle\bar\psi\psi\rangle =  c_0\Lambda_{QCD}^3 + 
\frac{c_1}{a^2}m_f + \frac{c_2}{a^3}e^{-L_S \alpha}.
\end{equation}
The last term in this equation can be estimated to be of the same size
as $\frac{1}{a^2}m_{res}$.
For the GeV case, this contributes an error 
to the value of $(12\langle\bar\psi\psi\rangle)^{1/3}$ quoted in 
Tab.~\ref{tab:pbp_phys} on the order of 20\%. 
This error is expected to decrease as $L_s \rightarrow \infty$.

\section{Conclusion}
Values for $\langle\bar\psi\psi\rangle$ are obtained from
different methods with different lattice scales.  Good
agreement is found in all but one case where a larger $L_s$
is required.

\section{Acknowledgment}
I thank the RBC members for providing the simulation results analyzed 
in this report and many useful suggestions,
especially Dr.C.Dawson for his help for NPR calculation,
and Prof.N.Christ for his help.


\begin{thebibliography}{2}
\bibitem{0007038} T. Blum, {\it et al. RBC}, hep-lat/0007038
\bibitem{RBC_02} Contributions of Y. Aoki, C. Dawson, J. Noaki and K. Orginos,
to Lattice Conference '02. Nucl.Phys.B.(Proc.Suppl) Vol.119.
\bibitem{RBC_03} Contributions of Y. Aoki, C. Dawson, J. Noaki and K. Orginos,
to Lattice Conference '03. 
\bibitem{NPR} T. Blum, {\it et al. RBC}, hep-lat/0102005
\end{thebibliography}
\end{document}